\documentstyle[twocolumn,prl,aps,epsfig]{revtex}

\begin{document}
\draft
\title{Field Driven Pairing State Phase Transition\\
in $d_{x^2-y^2}+id_{xy}$-Wave Superconductors}
\author{Bo Lei$^1$, S. A. Aruna$^1$, and Qiang-Hua Wang$^{1,2}$}
\address{$^1$Physics Department and National Laboratory of Solid State Microstructures,\\
Institute for Solid State Physics, Nanjing University, Nanjing 210093, China}
\address{$^2$Physics Department, University of California at Berkeley, CA94720}

\twocolumn[\hsize\textwidth\columnwidth\hsize\csname@twocolumnfalse\endcsname
\maketitle
\begin{abstract}
Within the framework of the Ginzburg-Landau theory for $%
d_{x^{2}-y^{2}}+id_{xy}$-wave superconductors, we discuss the pairing state
phase transition in the absence of the Zeeman coupling between
the Cooper pair orbital angular momentum and the magnetic field.
We find that above a temperature $T_{\ast }$, the pairing state in
a magnetic field is pure $d_{x^{2}-y^{2}}$-wave. However, below $T_{\ast }$,
the pairing state is $d_{x^{2}-y^{2}}+id_{xy}$-wave at
low fields, and it becomes pure $d_{x^{2}-y^{2}}$-wave at higher fields.
Between these pairing states there exists a field driven phase transition .
The transition field increases with decreasing temperature. In the
field-temperature phase diagram, the phase transition line is obtained
theoretically by a combined use of a variational method and the Virial
theorem. The analytical result is found to be in good agreement with
numerical simulation results of the Gingzburg-Landau equations. The validity
of the variational method is discussed. The difference to the case with
the Zeeman coupling is discussed, which may be utilized to the detection of
the Zeeman coupling. 
\end{abstract}

\pacs{74.60.Ec, 74.25.Dw, 74.20.De}
\date{Jan 3, 2000}

] 
\vskip2pc \narrowtext

The phase-sensitive experiment with a tricrystal superconducting ring
magnetometry\cite{Tsuei} demonstrated that in high temperature superconductors,
the dominant pairing channel is the $d_{x^{2}-y^{2}}$-wave one. However, it
is not yet clear whether there were some sub-dominant pairing channels.
It is even less clear what would be the symmetry of a sub-dominant pairing
channel if it existed at all. Such questions arise from 
a number of experiments, {\it e.g.}, the
observation of surface-induced broken time-reversal-symmetry ($T$ hereafter)
in YBCO tunnel junctions\cite{Covington}, the observation of fractional
vortices trapped in a boundary junction\cite{Kirtley}, and the abnormal
field dependence of the low temperature thermal conductivity $\kappa _{e}$
in BSCCO superconductors\cite{Krishana,Aubin}. These unusual phenomena can
not be adequately explained by the $d_{x^{2}-y^{2}}$-wave
pairing channel alone. Some sub-dominant channels (such as $s-$ or $d_{xy}$%
-wave ones) might have played a role in these phenomena. Thus it is interesting
to study theoretically the properties of the superconductors in the presence
of sub-dominant pairing channels.

As a model study, we consider the relevant singlet sub-dominant pairing
channel to be the $d_{xy}$-channel, which has been hotly discussed recently
in the context of the abnormal thermal conductivities in BSCCO
superconductors.\cite{Krishana,Aubin,Laughlin,Qhwang}. In this paper, using
a Ginzburg-Landau (GL) theory for $d_{x^{2}-y^{2}}+id_{xy}$-wave
superconductors,\cite{Qhwang} we discuss the pairing state phase transition
driven by the magnetic field. In the absence of the Zeeman coupling between 
the Cooper pair orbital angular momentum and the magnetic field (see 
below),\cite
{Qhwang} we find that above a temperature $T_{\ast }$, the pairing state in
a magnetic field is pure $d_{x^{2}-y^{2}}$-wave. However, below $T_{\ast }$,
the pairing state is $d_{x^{2}-y^{2}}+id_{xy}$-wave at
low fields, and it becomes pure $d_{x^{2}-y^{2}}$-wave at higher fields.
There exists a field driven phase transition between these pairing states.
The transition field increases with decreasing temperature. In the
field-temperature phase diagram, we are able to obtain the phase transition
line using a variational method and the Virial theorem. The analytical
result is in good agreement with that obtained from numerical simulation of
the Ginzburg-Landau equations.

The GL free energy of a $d_{x^{2}-y^{2}}+id_{xy}$-wave superconductor can be
obtained from the modified Bardeen-Cooper-Shrieffer (BCS) gap equation, the
Gor'kov theory or path integral formulation (in the weak coupling limit), 
\cite{Qhwang,Koyama,Wang}, 
\begin{eqnarray}
F &=&\int_{\Omega }\alpha _{D}|D|^{2}+\alpha _{D^{\prime }}|D^{\prime
}|^{2}+\Gamma \lbrack 3|D|^{4}/8+3|D^{\prime }|^{2}/8  \nonumber \\
&&+|D|^{2}|D^{\prime }|^{2}/4+(D^{\ast }D^{\prime }+{\rm c.c.})^{2}/8] 
\nonumber \\
&&+K\left[ |{\bf \Pi }D|^{2}+|{\bf \Pi }D^{\prime }|^{2})\right]
+\int_{\Omega }({\bf \nabla \times A})^{2}/8\pi ,  \label{Eq:Fenergy}
\end{eqnarray}
where $\int_{\Omega }$ denotes integration over the $ab$-plane, ${\bf \Pi }%
=-i{\bf \nabla }-2e{\bf A}/\hbar c$ is the gauge invariant gradient and $%
{\bf B=\nabla \times A}$ is the local induction (in the $z$-, or $c$-,
direction). 
Here $D$ and $D^{\prime }$ are order parameters in the $
d_{x^{2}-y^{2}}$- and $d_{xy}$-channels, respectively. $\alpha _{i}=N(0)\ln
T/T_{i}$ ($i=D,D^{\prime }$), where $N(0)$ is the normal state density of
states, and $T_{i}$ is the bare superconducting critical temperature in the $%
i$ channel. $K$ and $\Gamma $ are material-dependent 
parameters derivable from
microscopic theories \cite{Qhwang} and are assumed to be temperature
independent here for simplicity. We assume that $T_{D^{\prime }}\ll T_{D}$, 
{\it i.e.}, the dominant pairing channel is the $d_{x^{2}-y^{2}}$-wave
channel, as this would be consistent with high \label{stop}temperature
superconductors. In this paper, we do not consider the Zeeman coupling term
$F_z\propto -iB(D^{\ast }D^{\prime }-{\rm c.c.
})$ in the free energy.\cite{Qhwang,Koyama,Wang} 
On one hand, $F_z$ needs further
identification by a strong coupling microscopic theory. On the other hand, the
pairing states in a magnetic field in the absence of $F_z$ is
interesting in its own right and should be compared to that in the presence
of $F_z$, which will be discussed in the concluding section.

In order to see the relative importance of the various contributions to the
free energy and for later convenience, we rewrite the free energy in terms
of dimensionless quantities as, 
\begin{eqnarray}
F &=&E_{c}\int_{{\bf r}}\{-|d|^{2}-\alpha |d^{\prime
}|^{2}+(|d|^{4}+|d^{\prime }|^{4})/2+|d|^{2}|d^{\prime }|^{2}/3  \nonumber \\
&&+(d^{\ast }d^{\prime }+{\rm c.c})^{2}/6+|{\bf \pi }d|^{2}+|{\bf \pi }%
d^{\prime }|^{2}+\kappa ^{2}b^{2}\},  \label{freeg}
\end{eqnarray}
where $E_{c}={H_{c}}^{2}\xi ^{2}/4\pi $ with $H_{c}$ and $\xi $ being the
thermodynamic critical field and coherence length, respectively, in the
absence of the $d_{xy}$-wave channel. All quantities under the integration
symbol are now dimensionless: $\alpha =\alpha _{D^{^{\prime }}}/\alpha
_{D}=\ln (T/T_{D^{\prime }})/\ln (T/T_{D})$, $d=D/D_{0}$, $d^{\prime
}=D^{\prime }/D_{0}$, $r=R/\xi $, $a=A/A_{0}$, and $b=B/B_{0}$. Here $D_{0}=%
\sqrt{-4\alpha _{D}/3\Gamma }$ is the value of $D$ in the absence of the
order parameter $D^{^{\prime }}$ and the magnetic field. $\kappa $ is the
bare GL parameter in the absence of $D^{\prime },$ and $A_{0}=\Phi _{0}/2\pi
\xi $ where $\Phi _{0}$ is the flux quantum. $\ B_{0}=A_{0}/\xi $ is the
corresponding upper critical field. Finally, ${\bf \pi }=-i{\bf \nabla }-%
{\bf a}$ now denotes dimensionless gauge invariant gradient (i.e., ${\bf \pi 
}={\bf \Pi }\xi )$. 

The important difference between the above GL free energy Eq.(\ref{freeg})
and that for a $d+is$-wave superconductor lies in the fact that in the
former, we have no mixed gradient coupling between the two order parameters. 
\cite{Xu} Therefore, we can expect that the single vortex remains to be
circularly symmetric. This should be contrasted to the four-fold and
two-fold symmetric vortices in $d+is$ wave superconductors.\cite{Li} An even
more subtle consequence of the absence of the mixed gradient is that there
are three possible solutions,
namely: the pure $d$ solution, the pure $d^{\prime }$ solution, and the
mixed-wave solution. The physical solution of the system corresponds to that
with the lowest free energy. In this respect, it is not impossible that
there would be a phase transition between these solutions in the
field-temperature phase diagram. Indeed, this picture was conjectured
recently from the behaviors of the superconductor at zero field and the
upper critical field,\cite{Qhwang} which we summarize for completeness as
follows. (i) At zero field, it is easy to see that the two order parameters
develop a relative phase difference of \ $\pm \pi /2$ in order to minimize
the free energy. This $d_{x^{2}-y^{2}}\pm id_{xy}$-wave pairing state is $T$%
-breaking, which was recently argued to be relevant to the abnormal thermal
conductivity in BSCCO superconductors.\cite{Laughlin,Qhwang} The amplitudes
of the two order parameters are $|d|^{2}=3(3-\alpha )/8$ $\ $and $|d^{\prime
}|^{2}=3(3\alpha -1)/8$. Thus, $d_{xy}$-wave appears only when $\alpha >1/3,$
with a $T$-breaking transition point at $\alpha _{\ast }=1/3$. According to
the above definition of $\ \alpha $, this amounts to a transition
temperature $T_{\ast }=\sqrt{{T_{D}^{\prime }}^{3}/T_{D}}$. The zero-field $%
T $-breaking transition is a second-order phase transition, in that $%
d^{\prime }$ emerges continuously at $T\leq T_{\ast }$. (ii) From the
linearized GL equations, it was found that the upper critical field $%
b_{c2}=B_{c2}/B_{0}=1$ as long as $\alpha \leq 1$ (or $T>0K)$. Moreover, at
the upper critical field the eigen solution to the linearized GL equations
indicates that the order parameter $d^{\prime }$ vanishes identically,
although it would be nonzero at the same temperature ($<T_{\ast })$ but at
zero field. (iii) Combining both aspects we believe that there should be a
field-driven pairing-state phase transition at $T<T_{\ast }$ (or $\alpha
>\alpha _{\ast }$), where at zero and low fields, the pairing state 
is the $d_{x^{2}-y^{2}}+id_{xy}$-wave, while at higher fields,
it transforms to the pure $d_{x^{2}-y^{2}}$-wave. The mechanism is clear.
Since the two order parameters are not coupled by mixed gradient terms, 
they are frustrated in the vortex state: while
the coexistence of them lowers the homogeneous
energy at $\alpha >\alpha _{\ast }$, winding of both order parameters
due to vortices increases the kinetic energy. 
The competing energies should drive a
phase transition. Moreover, the transition field should increase with $%
\alpha $ as the gaining of homogeneous energy increases relatively. On the
other hand, at $T>T_{\ast }$ the pairing state is always the pure $%
d_{x^{2}-y^{2}}$-wave at any fields below the upper critical field. This
picture will be discussed in detail, in the proceeding section.

In our case, there are two order parameters. Even approximate solutions to
the GL equations, such as for the conventional superconductors,\cite{Tinkham}
are difficult to obtain. Thus we shall restrict ourselves to the variational
treatment of the system. The basic idea is as follows. We employ a reference
conventional system with one order parameter $\psi $, and set $d$ and $%
d^{\prime }$ to vary in space in the same manner as $\psi $ does, but with
different amplitudes which are to be optimized. Consequently, the behavior
of the present system is then approximately given by these amplitudes.

For our purpose, let us recapitulate some known results for the reference
system. The free energy is written as $F=E_{c}\int_{{\bf r}}\{-|\psi
|^{2}+|\psi |^{4}/2+|{\bf \pi }\psi |^{2}+\kappa ^{2}b^{2}\}.$ Two
identities follow immediately from the solution that minimizes this free
energy. Firstly, 
\begin{equation}
e_{2}+2e_{4}+e_{k}=0,  \label{identity1}
\end{equation}
where $e_{2}=\langle -|\psi |^{2}\rangle ,$ $e_{4}=\langle |\psi
|^{4}/2\rangle $ and $e_{k}=\langle |{\bf \pi }\psi |^{2}\rangle $. Here $%
\langle \cdot \rangle $ denotes spatial average. Second, $2{\kappa }%
^{2}h\langle b\rangle =e_{k}+2{\kappa }^{2}\langle b^{2}\rangle$, which is
known as the Virial theorem.\cite{Doria,Note}In the limit $\kappa \gg 1$,
the magnetic induction can be safely treated as being uniform at $h\gg
h_{c1} $ (with $h_{c1}$ being the dimensionless lower critical field), so
that an approximate Virial theorem can be written as
\begin{equation}
e_{k}\approx 2{\kappa }^{2}[h(b)-b]b,  \label{Eq:ek}
\end{equation}
where $h(b)$ is the magnetization curve which can be read off from the
textbook.\cite{Tinkham}

We now come back to the $d_{x^{2}-y^{2}}+id_{xy}$-wave superconductors.
As mentioned above, we set $d=\mu \psi $
and $d^{\prime }=\pm i\nu \psi $, with two real and
positive variational amplitudes $\mu $ and $\nu $. The $d^{\prime }$ order
parameter adopts a residual relative phase to $d$ in order to lower the free
energy Eq.(\ref{freeg}). Substitution of these ingredients into Eq.(2)
yields a free energy in terms of $\mu$, $\nu$, and $e_i$ ($i=2,4,k$). In
dimensionless form, the free energy density is
\begin{eqnarray}
f &=&\mu ^{2}e_{2}+\alpha \nu ^{2}e_{2}+(\mu ^{4}+\nu ^{4})e_{4} 
\nonumber \\
&&+2\mu ^{2}\nu ^{2}e_{4}/3+(\mu ^{2}+\nu ^{2})e_{k}+{\kappa }^{2}b^{2}.
\end{eqnarray}
Minimizing $f$ with respect to $\mu $ and $\nu $, we get: 
\begin{eqnarray}
&&\mu ^{2}=3[(\alpha -3)e_{2}-2e_{k}]/(16e_{4});  \label{dwave} \\
&&\nu ^{2}=3[(1-3\alpha )e_{2}-2e_{k}]/(16e_{4}).  \label{d1wave}
\end{eqnarray}

At low and intermediate fields, the magnetization curve in the reference
system in the limit $\kappa\gg 1$ is: $h=b+\ln (1/b)/2{\kappa }^{2}$
(in dimensionless form). \cite
{Tinkham} In combination with Eq.(\ref{Eq:ek}), 
we have $e_{k}=b\ln (1/b)$. On the
other hand, in the present field regime, the distance between two vortices
is much larger than the coherence length, hence we can assume $e_{2}\simeq
-1 $ and $e_{4}\simeq 1/2.$ This approximation would certainly violate Eq.(%
\ref{identity1}) up to the order of $\ e_{k}$ ($\ll 1$), but suffices to
yield an order of magnitude estimation of $\mu $ and $\nu $. Substituting
the approximate $e_i$'s into Eqs.(\ref{dwave}) and (\ref{d1wave}), we 
get $\mu ^{2}\approx
3(3/2-\alpha /2-b\ln 1/b)/4,$ and $\nu ^{2}\approx 3(-1/2+3\alpha /2-b\ln
1/b)/4$. We clearly see that $\mu ^{2}>0$ as long as $\alpha \leq 1$
({\it i.e.}, $d_{xy}$-channel is sub-dominant). This
means that the $d_{x^{2}-y^{2}}$-wave order parameter $d$ is always present in
the system. The $d_{xy}$-wave order parameter $d'$ is present provided 
that $\nu^2>0$. Thus the field driven phase transition is determined by 
$\nu ^{2}=0$, or
\begin{equation}
b\ln 1/b=(3\alpha -1)/2.  \label{linelow}
\end{equation}

At high fields closer to the upper critical field, the vortices are densely
distributed, hence the system can not be treated as above because the order
parameters are drastically suppressed by the magnetic field. Fortunately, in
this case we can resort to the high-field Abrikosov vortex lattice solution.
In the reference system one has $2e_{4}=\beta _{A}e_{2}^{2}$, where $\beta
_{A}$ is the Abrikosov constant. We combine this with Eq.(\ref{identity1}%
) to find $e_{2}=(\sqrt{1-4\beta _{A}e_{k}}-1)/(2\beta _{A})$. On the
other hand, the dimensionless magnetization curve in high fields is $%
h=b+(1-b)/(2{\kappa }^{2}\beta _{A})$,\cite{Tinkham} substitution of which
into Eq.(\ref{Eq:ek}) gives $e_{k}=b(1-b)/\beta _{A}$. Putting
the above together, we have in the high field regime: $\mu ^{2}=3[(\alpha
-3)e_{2}-2e_{k}]/(8\beta _{A}e_{2}^{2})$, and $\nu ^{2}=3[(1-3\alpha
)e_{2}-2e_{k}]/(8\beta _{A}e_{2}^{2})$, where both $e_{2}$ and $e_{k}$ can
be expressed as functions of $b$. Again the condition $\nu ^{2}=0$
determines the field-driven phase transition line: 
\begin{equation}
b=(1+\sqrt{9\alpha ^{2}-12\alpha +4})/2.  \label{linehigh}
\end{equation}

The phase transition lines Eqs.(\ref{linelow}) and (\ref{linehigh}) are
the main results of this work. They are plotted in Fig.1 (solid lines). 
Note that we have skipped the unphysical
portions of Eq.(\ref{linelow}) at high fields and Eq.(\ref{linehigh}) at low
fields. Also note that the dimensionless upper critical field is unity, but
the physical upper critical field $B_{0}$ depends on temperature. On the
other hand, $\alpha $ depends on temperature also (see the definition
above). The field regime between the two solid lines in Fig.1 can not be
determined theoretically by our approach, in that the order parameters are
already suppressed by the field so that the London approximation fails on
one hand, but the Abrikosov solution is still not reliable enough on the
other hand.

In order to check the accuracy of the variational method, we now obtain the
transition line by numerical simulation 
of the GL equations derived from Eq.(2). 
The simulation is performed in a unit cell of the
vortex lattice and the magnetic induction is treated as being uniform for
simplicity. The latter assumption is suitable at not too low fields and at $
\kappa \gg 1$, which is relevant to high-$T_{c}$ superconductors. The
simulation method is well documented in the literature.\cite{wang96} From
our simulation result, it is verified that the local relative phase
difference between the two order parameters is indeed $\pm \pi /2$. This
result provides a strong support to the validity of our analytical
treatment. The maximum amplitude of the order parameter $d^{\prime }$ in the
vortex solution, namely, $|d^{\prime }|_{\max }\equiv |D^{\prime }|_{\max
}/D_{0}$, as a function of $\alpha $ at various magnetic fields is plotted
in Fig.2, where we can clearly see that at a fixed field, $d^{\prime }$
drops to zero at a specific value of $\alpha $. This signals a field-driven
phase transition from the $d_{x^{2}-y^{2}}+id_{xy}$-wave pairing state
to the pure $d_{x^{2}-y^{2}}$-wave one (with vortices in the system). 
The transition
value of $\alpha $ decreases with decreasing magnetic field. The set of
transition points are plotted in Fig.1 (squares). In the high field regime,
the analytical result is in good agreement with the numerical result,
whereas in the low field regime, the result is only in qualitative
agreement. This is understandable from the fact that we have neglected the
vortex core energy and have adopted a crude approximation for the spatial
variation of the order parameters in our analytical treatment. In principle,
the two order parameters have different coherence lengths. They determine 
the length scales for the spatial variations of the order parameters.
Our approximation is equivalent to assign the same coherence 
length to both order parameters. The
good agreement at high fields is also understandable from the view that in
this case the vortices are strongly overlapped so that the length scale
approximation is not essential. In numerical simulations
we find some signs of abrupt drop of $d^{\prime }$ as a
function of $b$ or $\alpha $ in the parameter space. This may points to
a weakly first-order transition. However, our analytical result
is clearly a second order phase transition 
within the specified approximation.

Before closing, let us comment on the difference of our system to that
with a Zeeman coupling. In the latter case, it
was shown that a $d_{x^2-y^2}+id_{xy}$ pairing state is always induced
by vortices.\cite{Qhwang,Wang,Meirong} 
In other words, there is no further pairing state phase transition
in intermediate magnetic fields. Moreover, Balatsky\cite{Balatsky} 
recently argued that the Zeeman coupling enhances superconductivity near the
upper critical field. These distinct differences to the case discussed above
can be utilized for the detection of the Zeeman coupling. 

\acknowledgments{This work was supported by the National Natural Science
Foundation of China and the National Centre for Research and Development of 
Superconductivity of China. QHW was also supported by the Berkeley Scholars
Program financed by the Hutchison Whampoa Company, Hong Kong.}


\begin{figure}
\epsfxsize=8cm
\epsfbox{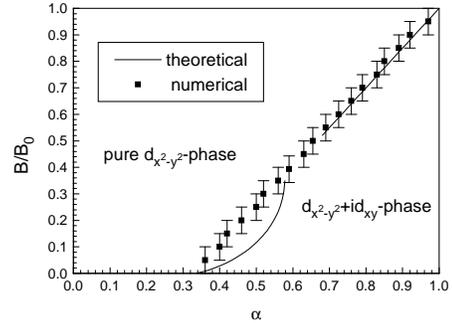}
\caption{Solid lines: analytical phase transition line; Squares: transition points 
extracted from numerical simulations. The error bars indicate the increment of the field
in the field-scanning.}
\end{figure}

\begin{figure}
\epsfxsize=8cm
\epsfbox{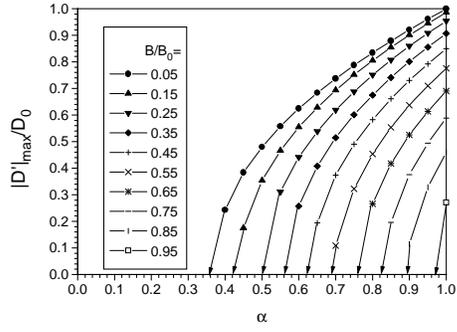}
\caption{The maximum of $|d'|$, {\it i.e.}, $|D'|_{\max}/D_0$
as a function of $\alpha$ at various fields. The arrows indicate the
estimated positions at which $d'$ drops to zero.}
\end{figure}


\begin{references}
\bibitem{Note}  The Virial theorem in full dimension reads: $\frac{1}{4\pi }%
HB=E_{k}+2E_{f}$ where $H$ is the applied magnetic field, $B$ is the average
induction field, $E_{k}$ is the kinetic energy of the superfluid, and
finally, $E_{f}$ is the magnetic energy in the GL free energy.\cite{Doria}
\end{references}
\end{document}